# Lexical Sorting Centrality to Distinguish Spreading Abilities of Nodes in Complex Networks under the Susceptible-Infectious-Recovered (SIR) Model

Aybike ŞİMŞEK [a,1]

[a] Düzce University Department of Computer Engineering, Düzce 81620, Turkey

Email: aybikesimsek@duzce.edu.tr

**Abstract**

Epidemic modeling in complex networks has become one of the latest topics in recent times. The Susceptible-Infectious-Recovered (SIR) model and its variants are often used for epidemic modeling. One important issue in epidemic modeling is the determination of the spreading ability of the nodes in the network. Thus, for example, can be detected in the early stages. In this study, we developed a centrality measure called Lexical Sorting Centrality **(**LSC) that distinguishes the spreading ability of the nodes. Using other centrality measures calculated for the nodes, LSC sorts the nodes in a way similar to alphabetical order. We conducted simulations on six datasets using SIR to evaluate the performance of LSC and compared LSC with degree centrality (DC), eigenvector centrality (EC), closeness centrality (CC), betweenness centrality (BC) and Gravitational Centrality (GC). Experimental results show that LSC distinguishes the spreading ability of the nodes more accurately, more decisively, and faster.

## 1. Introduction

Complex networks are very convenient tools for modeling the real world because many things in the real world connect together to form a complex network [1]. Complex networks have applications in many areas, including biology [2], social networks [3], ecology [4], and more. In addition, there are many practical benefits to determining the spreading capacities of the nodes in complex networks under certain propagation models. Some of these include: identification of nodes that can speed the spread of information in a social network [5], [6], detection of nodes that can minimize the spread of a rumor [7], [8], and revealing superspreaders that will play a role in the propagation of epidemics or computer viruses [9]. In this context, determining the spreading abilities of nodes in complex networks has attracted the attention of researchers [10]. Propagation models such as Susceptible-Infectious-Recovered (SIR), however, require heavy Monte-Carlo simulations. Therefore, researchers have turned to less costly methods that can indirectly determine the spreading abilities of nodes. Centrality measures are perhaps among the most frequently used of these methods. Many centrality measures have been developed and are still being developed for this purpose. Among the most basic of these are the degree, closeness, betweenness, eigenvector, and Katz centrality measures. The degree is a local measure of centrality that returns information on the number of nodes to which a node is directly connected. The closeness is a global measure of centrality calculated using the average distance of one node from other nodes (via the shortest paths). The betweenness is another measure of global centrality and gives information on the number of shortest paths (between all other node pairs) that pass through a node. According to the eigenvector, a node takes on the strength of its powerful adjacent node [11]. The Katz centrality has the same logic as that of the eigenvector. Eigenvector and Katz centrality measures are also global centrality measures. However, the Katz does not only take into account the strength of neighboring nodes. According to Katz centrality measure, a node makes use of the strength of the nodes connected to it by paths via the attenuation factor depending on the distance of the path [12].

---

[1] Corresponding author

Many new centrality measures have been developed other than the centrality measures given above. The general problem with these measures is that they cannot provide a solution to the balance between speed and quality. In this study, we developed a method we call Lexical Sorting Centrality (LSC). The LSC sorts the nodes according to their importance (effects) by using a combination of different centrality measures. In doing so, it does not use a predetermined closed-form solution, a factor which makes the LSC improvable. With this feature, LSC offers a new way to use different centrality measures in combination.

*Basic concept*

Different centrality measures can be calculated for nodes in complex networks. All measures calculate the importance (effect) of a node from its own perspective. However, a centrality measure does not yield the same performance in all complex networks. Therefore, the idea has emerged recently of combining the centrality measures [13]–[19]. Almost all of the studies, the details of which will be discussed in the next section, suggest combining the centrality measures with a closed formula. A merger in this way makes it difficult (if not impossible) to explain the logic of creating the formula when there is no natural phenomenon that inspired its creation. This also prevents the proposed methods from being improvable. In this study, we propose a way to use different centrality measures together without using a closed formula. We termed this method Lexical Sorting Centrality (LSC). Using multiple centrality measures, LSC sorts nodes in a manner similar to reverse lexical (alphabetical) order. For example, three different centrality measures ($C_1$, $C_2$, and $C_3$) are calculated for a six-node graph. For convenience, the precision of the centrality measures is taken as one decimal place. Each centrality measure here is like a letter and the node centrality measure array can be thought of as a word. For example, in Figure 1, the letters formed in node 0 are

0.2, 0.8, and 0.3. Figure 1.(a) shows the initial state. Here, the nodes are ordered according to their own numbers. In situation Figure 1.(b) the nodes are sorted from large to small according to $C_1$ measure. Here, the $C_1$ values of nodes 4 and 5 and nodes 0 and 2 are the same. Therefore, it is not possible to say which of these nodes is more important than the other by looking only at the $C_1$ measure. Afterwards, only the nodes with the same $C_1$ values (i.e., nodes 4 and 5 and nodes 0 and 2) are sorted from large to small according to $C_2$ values to arrive at the situation in Figure 1.(c). The $C_1$ and $C_2$ values of nodes 0 and 2 are the same. Finally, the nodes with the same $C_1$ and $C_2$ values (nodes 0 and 2) are ranked among themselves according to the $C_3$ value to reach the final state in Figure 1.(d). The final arrangement of the nodes is 5-4-1-2-0-3. If we consider each centrality measure as a letter here, this order is exactly that of a reverse lexical. If all centrality measures calculated for any two nodes are the same, the order of those nodes is not changed.

| Node | $C_1$ | $C_2$ | $C_3$ | Node | $C_1$ | $C_2$ | $C_3$ | Node | $C_1$ | $C_2$ | $C_3$ |
|---|---|---|---|---|---|---|---|---|---|---|---|
| 0 | 0.2 | 0.8 | 0.3 | 4 | **0.7** | 0.5 | 0.1 | 5 | **0.7** | **0.6** | 0.7 |
| 1 | 0.5 | 0.3 | 0.5 | 5 | **0.7** | 0.6 | 0.7 | 4 | **0.7** | **0.5** | 0.1 |
| 2 | 0.2 | 0.8 | 0.4 | 1 | **0.5** | 0.3 | 0.5 | 1 | **0.5** | **0.3** | 0.5 |
| 3 | 0.1 | 0.4 | 0.8 | 0 | **0.2** | 0. | 0.3 | 0 | **0.2** | **0.8** | 0.3 |
| 4 | 0.7 | 0.5 | 0.1 | 2 | **0.2** | 0.8 | 0.4 | 2 | **0.2** | **0.8** | 0.4 |
| 5 | 0.7 | 0.6 | 0.7 | 3 | **0.1** | 0.4 | 0.8 | 3 | **0.1** | **0.4** | 0.8 |
| (a) | | | | (b) | | | | (c) | | | |

| Node | $C_1$ | $C_2$ | $C_3$ |
|---|---|---|---|
| 5 | **0.7** | **0.6** | **0.7** |
| 4 | **0.7** | **0.5** | **0.1** |
| 1 | **0.5** | **0.3** | **0.** |
| 2 | **0.2** | **0.8** | **0.4** |
| 0 | **0.2** | **0.8** | **0.3** |
| 3 | **0.1** | **0.4** | **0.8** |

(d)

**Fig. 1.** Sorting nodes in a manner similar to alphabetical order using multiple centrality measures.

## 2. Related Work

Centrality measures such as degree [20], closeness [21], betweenness [22], eigenvector [11], Katz [12], and PageRank [23] are well-known centrality measures used to determine the importance of nodes in complex networks. In this section, we will discuss the more recent centrality measures and studies that use more than one measure together.

Andrade and Rêgo developed a centrality measure called p-means [24]. The p-means combines degree, closeness, harmonic, and eccentricity centrality measures parametrically into one formula. By changing the coefficient in the formula of p-means, the p-means behaves like one of these measures (or a combination of them). Using the degree of nodes, Zhao et al. developed the two sub-measure of self-importance and global importance and by multiplying them, proposed a composite centrality measure demonstrating the importance of the nodes [25]. By using a coefficient for self-importance and global importance in their formula, they ensured that the balance could be changed towards one of these two sub-measure. They compared their proposed method with well-known centrality measures such as degree, closeness, betweenness, and PageRank on six real networks under the SIR propagation model and achieved competitive results. Alshahrani et al. developed two algorithms called MinCDegKatz d-hops and MaxCDegKatz d-hops [17]. These algorithms combine degree and Katz centrality measures, taking into account the local and general strength of the nodes. They compared their proposed algorithms to various competing algorithms under Independent Cascade (IC) and Linear Threshold (LT) propagation models on four real networks. Yang et al. developed a new centrality measure called DCC [19], which combines degree and clustering coefficient measure in a closed formula. Just like in the measure proposed by Zhao et al. [25], by using a coefficient for the sub-components in their formula, they ensured that the balance could be changed towards one of these two sub- measures. They compared their proposed algorithms to different competing algorithms under the Susceptible – Infected (SI) propagation model on four real networks. Şimşek and Meyerhenke developed many new centrality measures using well-known centrality measures such as degree, closeness, and eigenvector [26]. The measures they developed are based on multiplying well-known measures with various coefficients and subjecting them to different arithmetic operations. They compared their proposed new measures with competing algorithms under the IC propagation model on 50 real networks.

In summary, many of the studies in the literature are based on combining multiple measures with a closed formula and making the formula adjustable by assigning a coefficient to each sub-measure (e.g., $factor_1 \bullet Centrality\ Measure_1 + factor_2 \bullet Centrality\ Measure_2$). The first problem here is to determine the values of the coefficients. Most of the studies empirically give the coefficients a value of "1". Another problem is the arithmetic operations to which the sub-measure in the developed compound measures are subjected. It is difficult to explain the purpose of multiplying or summing the two sub-measures with each other.

## 3. Distinguishing Spreading Abilities of Nodes Using LSC

### 3.1. Preliminary information

As mentioned in the introduction, LSC sorts nodes in a way similar to reverse lexical order using multiple centrality measures. In this study, we selected the degree, eigenvector, and closeness measures. We also used Susceptible-Infectious-Recovered (SIR) as the propagation model. First, let us discuss these measures and the SIR model.

Let $G = (V, E)$ be an undirected unweighted graph (network). Here, $V$ is the set of nodes (vertices), and $E$ is the set of edges (links).

**Definition 1** (*Degree Centrality*): Degree centrality (DC) is calculated by dividing the degree of the node by the total number of nodes in the graph minus 1.

$$DC(i) = \frac{degree(i)}{|V| - 1} \tag{1}$$

Here, $i \in V$.

**Definition 2** (*Eigenvector Centrality*): The eigenvector centrality (EC) value of a node is obtained by dividing the sum of the EC values of neighboring nodes by a constant.

Let $A = (a_{ij})$ be the adjacency matrix of $G$; if $i$ and $j$ are neighbors, $a_{ij} = 1$; otherwise, $a_{ij} = 0$.

$$EC(i) = \frac{1}{\lambda} \sum_{j \in V} a_{ij} EC(j) \tag{2}$$

Here, $\lambda$ is a constant.

**Definition 3** (*Closeness Centrality*): The closeness centrality (CC) value of a node is calculated using the average distance (over the shortest paths) of the node to all other nodes.

$$CC(i) = \frac{|V|}{\sum_{j \in V - \{i\}} sp(j, i)} \tag{3}$$

Here, $sp(\bullet)$ is the shortest path between nodes $i$ and $j$.

**Definition 4** (*Susceptible-Infectious-Recovered Model*): Susceptible-Infectious-Recovered (SIR) is a well-known epidemic model. Although it is a population-based model, because of its popularity in recent years, it is beginning to be applied to network structures [27]. The SIR model is defined by two parameters: $\beta$; the rate at which the sensitive nodes are infected by their already infected neighbors; and $\gamma$; the recovery rate of infected nodes. Initially, all nodes are susceptible. One or more nodes on the network are first infected (the nodes that bring the disease to the network). These are often referred to as core nodes. Starting from these nodes, in step $t$ (time frame), the infection spreads over the network and becomes an epidemic, or otherwise, it is contained before it becomes an epidemic. When the infection becomes an epidemic, the number of nodes infected first and their location (significance) on the network depends on the network topology (i.e., density), $\beta$ and $\gamma$.

### 3.2. LSC

The LSC sorts nodes in a way similar to reverse lexical order, using multiple centrality measures, as explained in the Introduction section. In this study, we used DC, EC, and CC measures. First, DC, EC, and CC are calculated for all nodes. Thus, a node and its calculated measures can be written as follows:

$$[node \quad DC \quad EC \quad CC]_{1\times 4}.$$

When this is done for all nodes, a matrix is produced, as shown in Equation (4). We named this the Ranking Matrix (RM).

$$\mathbf{RM} = \begin{bmatrix} node_0 & DC_0 & EC_0 & CC_0 \\ node_1 & DC_1 & EC_1 & CC_1 \\ \cdot & \cdot & \cdot & \cdot \\ \cdot & \cdot & \cdot & \cdot \\ \cdot & \cdot & \cdot & \cdot \\ node_{n-1} & DC_{n-1} & EC_{n-1} & CC_{n-1} \end{bmatrix}_{n\times 4} \quad (4)$$

Here, $n = |V|$.

The reverse lexical sorting was then performed as follows:

The RM is first sorted by the *DC* column. Thus, the rows with the same *DC* values (if any) will come one after another. These rows are sorted according to their *EC* values. Rows with the same *EC* values (if any) will come one after another. These rows are sorted among themselves according to the *CC* values. Thus, the ranking is completed. If we consider each centrality measure as a letter, then each node can be thought of as a three-letter word. Therefore, the method is called Lexical Sorting. The LSC can be improved for more than one desired centrality measure. We used three centrality measures here.

The order of the selected measures is an important point. The centrality measure which is at the forefront in the **RM** has priority, just like starting from the first letters of the words in alphabetical order. The DC is an important local centrality measure. If all nodes have the same probability of infection ($\beta$), a high-degree node has the chance to infect more nodes at once. We used the EC as a secondary measure as it incorporates the powers of the neighbors into the account. If the probability of a node infecting a neighbor is β, the probability of infecting its neighbor's neighbor is $\beta^2$ (usually $\beta$ is chosen as much smaller than 1). Since the probability will be $\beta \gg \beta^2$, we gave DC priority over EC. Finally, we used CC because it provides information about the location of the node on the network.

Another important point is the decimal precision of the measure. The selected decimal digit precision will change the order. Let us consider the nodes in Equation (5). Since the DC value of $node_0$ is greater than the DC value of $node_1$, the LSC positions $node_0$ first and $node_1$ second. However, if we take the precision of the digits to only two places after the decimal point, the DC values of $node_0$ and $node_1$ will be the same, as seen in Equation (6). In this case, the LSC positions $node_1$ first and $node_0$ second because LSC sorts the nodes by EC values (and since here the EC values of the two nodes are different). The order will be as in Equation (6). In this study, we choose the decimal precision to five places. Thus, we preserved the ranking of the priority measures as much as possible.

$$\mathbf{RM} = \begin{bmatrix} node_0 & 0.76525 & 0.05963 & 0.15423 \\ node_1 & 0.76234 & 0.06421 & 0.24563 \end{bmatrix} \quad (5)$$

$$\mathbf{RM} = \begin{bmatrix} node_1 & 0.76 & 0.06 & 0.24 \\ node_0 & 0.76 & 0.05 & 0.15 \end{bmatrix} \quad (6)$$

## 4. Experiments

In order to evaluate the performance of the LSC, we selected five competing centrality measures and conducted experiments on 1 synthetic and 5 real world network datasets. First, let us discuss the competitive measures and datasets.

### 4.1. Centrality measures

DC (*Degree Centrality*) [20]: The DC is calculated by dividing the degree of the node by the total number of nodes in the graph minus one. The DC, one of the main centrality measures, is a local measure.

EC (*Eigenvector Centrality*) [11]: The EC is obtained by dividing the sum of EC values of the nodes to which the EC node is directly connected (i.e., neighbors) by one constant. To do this, each node is initially assigned a specific EC value.

CC (*Closeness Centrality*) [21]: The CC is calculated using the average distance (over the shortest paths) of the CC node to all other nodes. The CC value of the node that is closest to all other nodes is the highest.

BC (*Betweenness Centrality*) [22]: The BC provides information on the number of times a node can intersect the shortest paths among all other node pairs It can be said that the BC values of the nodes that serve as bridges will be high.

GC (*Gravitational Centrality*) [18]: The GC is a recent centrality measure whose development was inspired by Newton's gravitational formula. In the GC, the k-shell values of the nodes replace the mass in Newton's formula. It uses the length of the shortest path between nodes, rather than the distance between masses. Its formula is as follows:

$$GC_i = \frac{ks_i \times ks_j}{\sum_{j \in N} sp(j,i)} \quad (7)$$

Here, $sp(\bullet)$ is the shortest path between nodes $i$ and $j$; $N$ is the set of 3-hop neighbors of node $i$.

The GC was chosen as a competitor because like LSC, it can use different centrality measures (e.g., k-shell here).

### 4.2. Datasets

We used 1 synthetic and 5 real world networks for the experimental studies. The properties of the networks are given in Table 1.

*Barabasi-Albert*: This synthetically-created scale-free network includes 1000 nodes and 9900 edges [28].

*Karate*: This network consists of 34 nodes and 78 edges. The nodes denote members of the club, and the edges denote the friendship between members [29]. This dataset is taken from http://konect.uni-koblenz.de/publications.

*Email-Enron*: This Email network consists of 143 nodes and 623 edges [30]. This dataset is taken from http://networkrepository.com.

*Email-Univ*: This network consists of 1133 nodes and 5452 edges [31]. This dataset is taken from http://konect.uni-koblenz.de/publications.

*CS-PhD*: This network consists of 1882 nodes and 1740 edges [32]. This dataset is taken from http://networkrepository.com.

*Ia-reality*: This network consists of 6809 nodes and 7680 edges [33]. This dataset is taken from http://networkrepository.com.

**Table 1.** Network dataset features

| Dataset | $|V|$ | $|E|$ | $\langle K \rangle$ | $K_{max}$ | Density |
| --- | --- | --- | --- | --- | --- |
| Barabasi-Albert | 1000 | 9900 | 19.8 | 198 | 0.0198198 |
| Karate | 34 | 78 | 4.588 | 17 | 0.1390374 |
| Email-Enron | 143 | 623 | 8 | 42 | 0.0613612 |
| Email-Univ | 1133 | 5452 | 9.62 | 71 | 0.0085002 |
| CS-PhD | 1882 | 1740 | 1.849 | 46 | 0.0009830 |
| Ia-reality | 6809 | 7680 | 2.256 | 261 | 0.0009830 |

### 4.3. Evaluation of the ranking performance of centrality measures

First, we evaluated the ranking performance of the centrality measures under the SIR model. For this, we used the Kendall *tau* correlation coefficient commonly used in the literature [34]. Let $(a_i, b_i)$ and $(a_j, b_j)$ be tuples of joint A and B ranking lists. If $a_i > a_j$ and $b_i > b_j$ or $a_i < a_j$ and $b_i < b_j$, then the tuples are concordant. If $a_i > a_j$ and $b_i < b_j$ or $a_i < a_j$ and $b_i > b_j$, then the tuples are discordant. If $a_i = a_j$ or $b_i = b_j$, then the tuples are neither concordant nor discordant. Finally, *tau* is defined as in Equation (8).

$$tau = \frac{N_c - N_d}{0.5N(N-1)} \quad (8)$$

Here, $N_c$ is the number of concordant pairs, $N_d$ is the number of discordant pairs, and $N$ is the number of all combinations. Positive $tau$ values indicate a positive correlation, and negative $tau$ values indicate a negative correlation. The ranking performances of LSC and other centrality measures under the SIR model are shown in Figure 2. In the simulations, the infection rate for small or less dense networks was $\beta = 0.1$ and for larger or more dense networks, $\beta = 0.01$. The recovery rate for all simulations was taken as $\gamma = 1$. Since it is difficult to differentiate the spreading abilities of nodes for large $\beta$ values, the $\beta$ value was chosen according to the scale of the network [35]. The simulations continued until there were no infected nodes on the network. The SIR score of a node is the total number of recovered nodes at the end of the simulation when that node is selected as the sole seed. All SIR simulations in this study were repeated 1000 times and their averages were used. NetworkX was used for the network operations [36].

The LSC performed the best in three datasets. Its performance in other datasets is very close to the benchmark that gave the best performance. Thus, the LSC was shown to yield good and stable results.

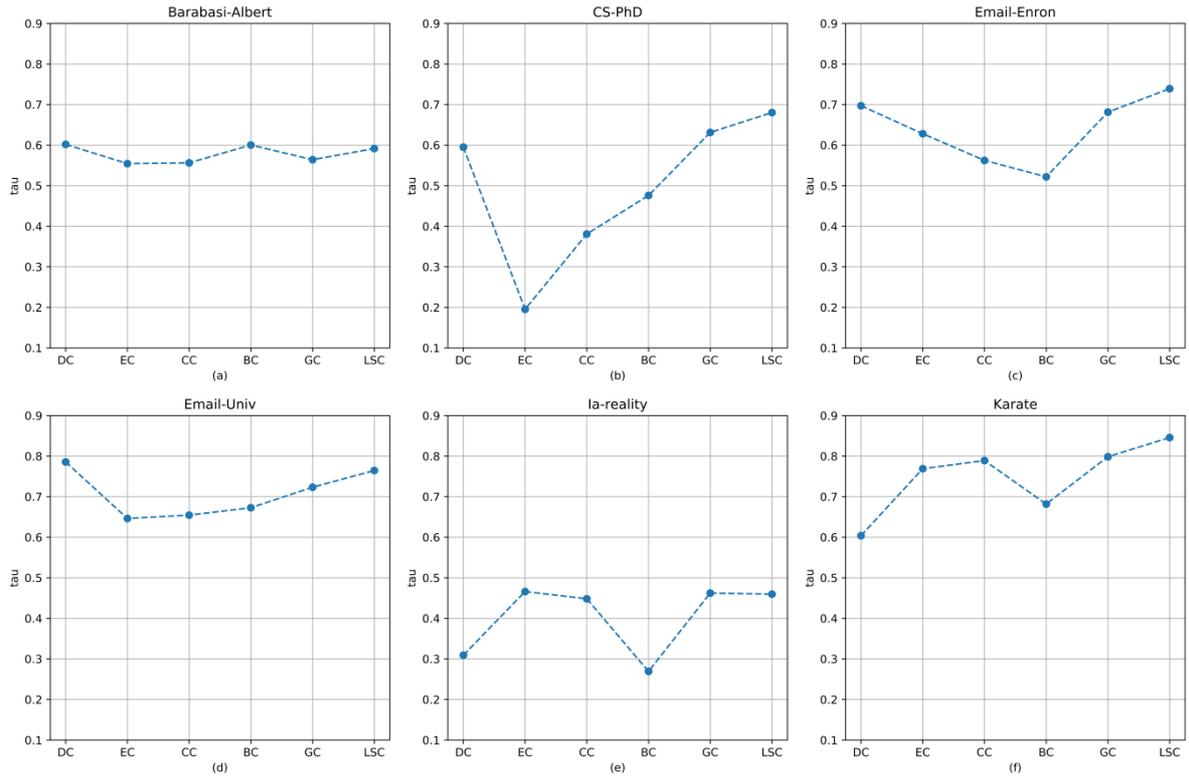

**Fig. 2.** Kendall $tau$ correlation coefficient values of different centrality measures. Infection rate: $\beta = 0.1$ for (b), (d), and (f); $\beta = 0.01$ for (a), (c), and (e). Recovery rate: $\gamma = 1$ for all experiments.

In addition, graphics of the ranking lists created by the centrality measures vs. the SIR scores are shown in Figure 3. A node with a lower index value is expected to have a higher SIR score. Therefore, a decrease in the SIR score as the index increases indicates the success of the centrality measure. In Figure 3, the graphics created by the LSC are for the most part smoother compared to those created by the other measures.

Additionally, when evaluating the performances of the centrality measures, the spreading abilities of the nodes determined as $top - x$ by the centrality measures were evaluated. To this purpose, firstly, in Table 2, we have shown the ranking lists of the centrality measures and the number of matching nodes in the top 5% of the ranking list of the SIR simulation. For example, the table shows that 5% of the number of nodes of the Barabasi-Albert network is 50. According to the SIR scores, 42 of the nodes that entered the top 50 when ranked from large to small were also in the top 50 of the LSC ranking list. LSC yielded the best results on all networks for both.

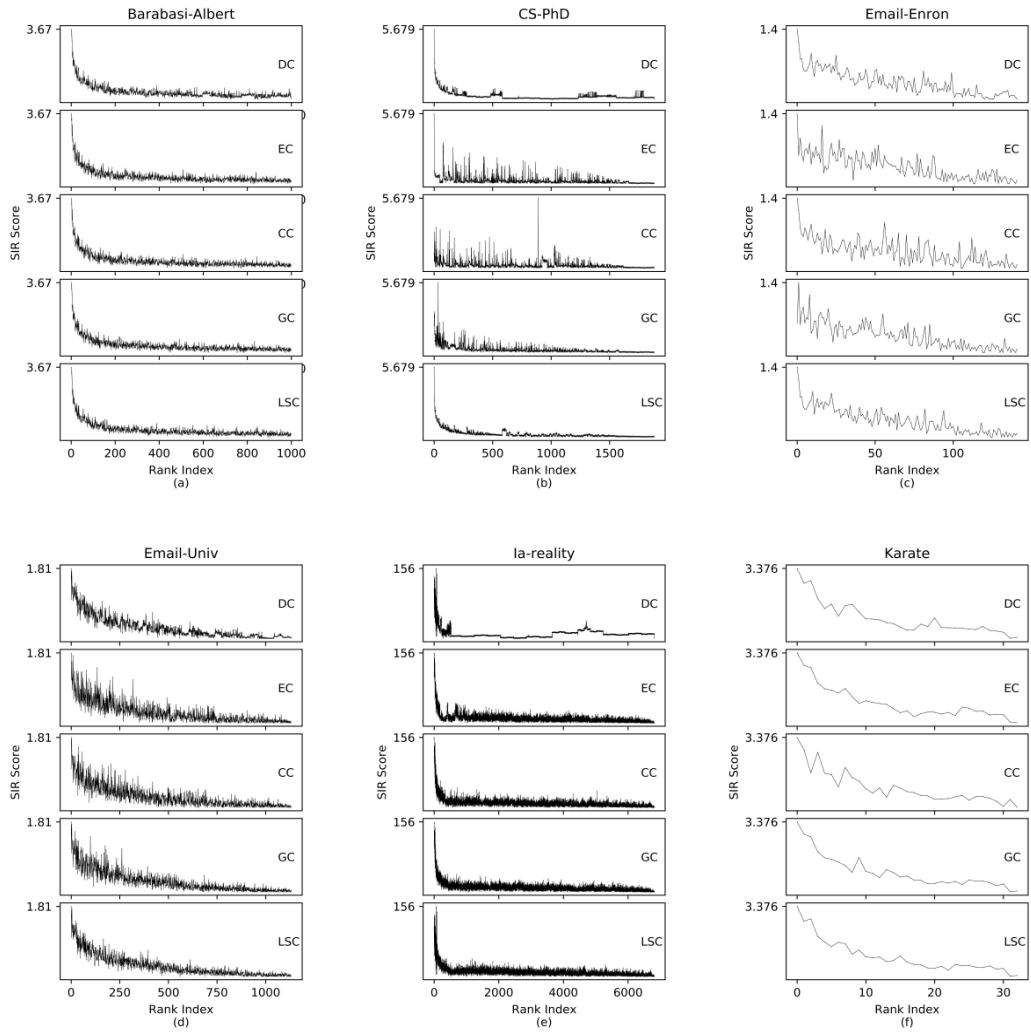

**Fig. 3.** SIR score trends of nodes sorted from large to small according to different centrality measures.

**Table 2.** Number of matching nodes in the top 5% of the ranking list of the centrality measures and the ranking list of the SIR simulation.

|  | DC | EC | CC | BC | GC | LSC |
|---|---|---|---|---|---|---|
| Barabasi-Albert | 40 | 42 | 42 | 41 | 42 | 42 |
| CS-PhD | 77 | 15 | 27 | 46 | 53 | 80 |
| Email-Enron | 4 | 2 | 4 | 4 | 3 | 4 |
| Email-Univ | 41 | 29 | 35 | 36 | 36 | 41 |
| Ia-reality | 201 | 166 | 245 | 199 | 245 | 247 |
| Karate | 1 | 1 | 1 | 1 | 1 | 1 |

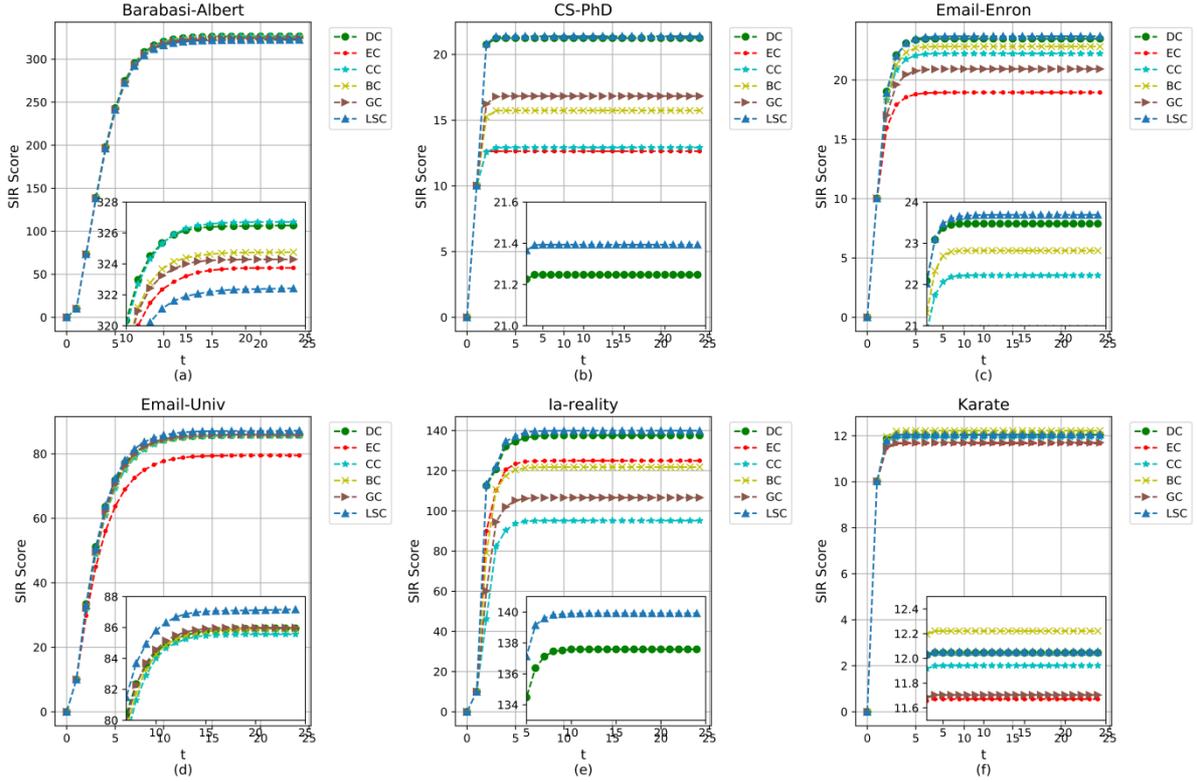

**Fig. 4.** SIR scores of the nodes in the top 5% determined by different centrality measures. ($t = 25, \beta = 0.05$)

Finally, we examined how the SIR scores (created according to the core of the $top - 10$ nodes determined by the centrality measures) changed over time. Trials were conducted for different $\beta$ values; however, only the result of $\beta = 0.05$ is given because other results were similar and we wanted to save space. The trials yielded $t = 25$. The trends are shown in Figure 4. The rapid increase of the curves in the graphs indicates that the nodes selected by the relevant centrality measure maximized the spread in a short time. According to the graphics, the nodes chosen by the LSC increased the spread to the highest level in a short time. In addition, it can be seen in the enlarged inserts in the graphics that in four datasets the nodes selected by the LSC have a higher spreading capacity.

### 4.4. Operating speed of LSC and GC

In addition to the performance of a centrality measure, its speed is also important. Because of this, we compared the calculation times of the LSC and GC on 6 datasets (Table 3). We ran the calculations 1000 times on a computer with an Intel i7 2.8 GHz processor and 16 GB of RAM and averaged the runtimes.

**Table 3.** Calculated times (s) of LSC and GC for six datasets.

|     | Barabasi-Albert | CS-PhD | Email-Enron | Email-Univ | Ia-reality | Karate |
| --- | --- | --- | --- | --- | --- | --- |
| LSC | 7.476 | 3.045 | 0.113 | 6.068 | 121.858 | 0.065 |
| GC | 51.467 | 6.011 | 0.367 | 11.351 | 101.591 | 0.005 |

In four of the six datasets the LSC was calculated as faster than the GC. The crucial time-consuming factor for the GC is the calculation of each node for its 3-hop neighbors. In dense networks, 3-hop neighbors make up a large part of the network. On the other hand, the chief time-consuming element of the LSC is the calculation of the centrality measures it uses. Once the centrality measures are calculated, it performs only the sorting process. When different centrality measures are used in the LSC, this leads to significant changes in the operating time. Also, a number of studies have been carried out with the aim of achieving faster calculation of centrality measures [37]. By using these methods, the operating time of the LSC can also be reduced.

## 5. Discussion and Conclusions

In this study, we proposed a new centrality measure for complex networks that ranks the nodes according to their spreading abilities under the SIR model. In a manner similar to lexical sorting, LSC sorts nodes according to multiple centrality measures. With this feature, it is not just a new centrality measures, but also a framework for creating new centrality measures. Our detailed simulations have demonstrated that LSC performs better than well-known centrality measures such as degree, closeness, eigenvector, and betweenness and the state-of-the-art GC measure. Future studies might consider using different centrality measures in different orders and a different decimal digit precision in LSC.

**Biography**

Aybike ŞİMŞEK received her BS degree from the Selçuk University Department of Computer Engineering in 2003, her MS degree from the Gazi University Department of Computer Engineering in 2010, and her PhD from the Düzce University Department of Electrical-Electronics and Computer Engineering in 2018. She was a postdoctoral researcher in the Department of Computer Science at the Humboldt University of Berlin from August 2018 to July 2019. She has been working as an assistant professor in the Department of Computer Engineering at Düzce University since January, 2020. Her current research interests include social network analysis, complex networks, and epidemic modeling.